\documentclass[twocolumn,showpacs,preprintnumbers,amsmath,amssymb]{revtex4}

\usepackage{graphicx}
\usepackage{epsfig}
\usepackage{dcolumn}
\usepackage{bm}

\begin{document}

\title{Predicted signatures of the intrinsic spin Hall effect in closed systems}

\author{Manuel Val\'{\i}n-Rodr\'{\i}guez\footnote{email: mvalin@iaqse.caib.es}}
\affiliation{%
Conselleria d'Educaci\'o i Cultura, Govern de les Illes Balears.
07004 Palma de Mallorca, Spain
}%

\date{\today}

\begin{abstract}
We study a two-dimensional electron system in the presence of spin-orbit interaction.
It is shown analytically that the spin-orbit interaction acts as a transversal effective electric field,
whose orientation depends on the sign of the $\hat{z}$-axis spin projection. This effect
doesn't require any driving electricl field and is inherent to the spin-orbit interactions
present in semiconductor materials. Therefore, it should manifest in both closed and open systems.
It is proposed an experiment to observe the intrinsic spin Hall effect in the far infrared absorption of 
an asymmetric semiconductor nanostructure.
\end{abstract}


\maketitle

In recent years, the study of spin related phenomena has become one of the most active 
fields of condensed matter physics. This interest is motivated by the possibility of a new generation
 of devices with more versatile functionality and lower power consumption \cite{zutic,fab,wolf} as well 
as for the fundamental physics involved. In particular, the spin Hall effect constitutes a paradigmatic 
example from both perspectives: applied and fundamental.

This effect involves the generation of a transverse spin current from a charge current through
a spin-orbit interaction mechanism, thus providing a tool to generate spin currents by electrical means 
which is an essential ingredient for spintronic applications. In this sense, the reciprocal effect is also 
relevant because it provides an electrical method to detect spin currents. Known as the inverse spin Hall 
effect, implies that a pure spin current also generates a transverse charge current in the presence 
of spin-orbit interaction \cite{hirsch}.

Since its original proposal in 1971 \cite{perel}, several papers addressed this effect \cite{hirsch,zhang,murakami,sinova1}
 and its first experimental evidences \cite{kato,wunder} triggered an intensive theoretical
\cite{nikolic1,halperin,shen,schliemann2,zhang2,rashba,nikolic2,sheng,kane,berne,sherman,garlid}
 and experimental \cite{invsh,sih,zhao,stern} study of the effect. 

The original proposal of the spin Hall effect relied on the spin-dependent scattering between electrons 
and impurities as the underlying physical mechanism to induce the transverse spin current. At present, this is known
 as the extrinsic spin Hall effect \cite{perel,hirsch,zhang}.
In contrast, when the mechanism responsible for the effect is the built-in spin-orbit interaction,
it is known as the intrinsic spin Hall effect which is of relevance in nanoscale systems where the transport is
 essentially ballistic.

In the pioneering work on the intrinsic spin Hall effect \cite{sinova1,murakami}, the transverse spin current originates
from the combined effect of spin-orbit interaction and a driving electric field. However,
calculations for ballistic nanostructures suggest that the driving electric field is not an essential
ingredient to induce the transversal spin separation \cite{nikolic1} which is intimately related to the spin-orbit interaction.
This is consistent with the recent experiments of Werake {\em et al.} \cite{invsh} where, by means of laser pumping, 
valence band electrons were excited to the conduction one of a sample with spin-orbit interaction  and left to evolve freely.
The dynamics of the electrons were monitored showing an unambigous proof of the inverse intrinsic spin Hall effect.

The spin-orbit coupling stems from the relativistic correction to the electronic motion.
In semiconductors, the built-in electric fields are felt by the electrons in their intrinsic
reference frame as a spin-dependent magnetic field. Depending on the origin of the electric
field, two sources of spin-orbit coupling are distinguished: the lack of inversion symmetry
in a bulk semiconductor originates the so-called Dresselhaus term \cite{dress}, whereas the asymmetry
of the confining potential corresponding to a semiconductor heterostructure is the
responsible of the Bychkov-Rashba interaction \cite{rash}. For simplicity, we consider first the effect of
the Bychkov-Rashba interaction alone.

The effective mass Hamiltonian for a ballistic two-dimensional electron system is given by

\begin{equation}
\label{horig}
{\cal H}=\frac{p_x^2+p_y^2}{2m^*}+V(x,y)+\frac{\alpha}{\hbar}\left(p_y\sigma_x-p_x\sigma_y\right),
\end{equation}

where $m^*$ represents de conduction band effective mass, $\alpha$ is the strength of Bychkov-Rashba spin-orbit
interaction and $V(x,y)$ is a generic confining potential that determines the geometry of the system.
 Following the approach of Aleiner and Fal'ko \cite{alei}, this Hamiltonian
can be diagonalized in spin space up to second order in the Bychkov-Rashba parameter '$\alpha$' by
means of an unitary transformation
\begin{equation}
\label{uspin}
{\cal U}_s=\exp\left\{-i\alpha\frac{m^*}{\hbar^2}\left(y\sigma_x-x\sigma_y\right)\right\}.
\end{equation}

The Hamiltonian in the transformed reference frame reads ($\tilde{\cal H}={\cal U}_s^{\dagger}{\cal H}{\cal U}_s$) ,

\begin{equation}
\label{transham}
\tilde{\cal H} =\frac{p_x^2+p_y^2}{2m^*}+V(x,y)-\alpha^2\frac{m^*}{\hbar^3}L_z\sigma_z -\frac{\alpha^2 m^*}{\hbar^2}+o(\alpha^3).
\end{equation}

The terms $o(\alpha^3)$ can be neglected as long as the effect of the spin-orbit interaction is much smaller than the tyipical orbital level spacing.
Since the effect of the spin-orbit interaction is tipically a small fraction of a meV, we can safely use this model for systems characterized by a level
spacing of the order of the meV or higher. This corresponds to systems with a typical size ranging from a few nanometers to a fraction of a micrometer.

The corresponding eigenstates, $\tilde{\phi}_i(\vec{r})\chi_s$, have well-defined spin orientation perpendicular to the plane ($\chi_\uparrow =(1,0)$ and $\chi_\downarrow =(0,1)$ in the $\sigma_z$-basis).
When transformed back to the original frame $\phi_i(\vec{r},\eta)= {\cal U}_s \tilde{\phi}_i(\vec{r})\chi_s$, these states become spinors, that, in a circularly symmetric system, 
recover the total angular momentum symmetry ($J_z$) as expected for the Bychkov-Rashba spin-orbit interaction.

The spin Hall effect is characterized by the generation of a transverse spin current from a charge
current. In a microscopic picture, this means that the electrons with a net motion in given
direction undergo a selective transverse deflection depending on the orientation of its spin.
To get in this context, a second unitary transformation is applied to the Hamiltonian

\begin{equation}
\label{uspin}
{\cal U}_d=\exp\left[-i q\left(\hat{n}\cdot\vec{r}\right)\right].
\end{equation}

This transformation has a clear physical interpretation, since it corresponds to a translation in the momentum space. The parameter
'$q$' represents the amplitude of the translation and the unitary vector $\hat{n}=(\cos\theta,\sin\theta)$ its direction,
 being $\theta$ the angle with respect to the x-axis. Physically, this transformation can model the effect of a time-dependent
electric field which modifies the momentum of the electrons in a given direction. The use of time-dependent electric fields to detect
the spin Hall effect has already been proposed in the context of the two-dimensional electron gas \cite{martin}. Consider an ultrafast pulsed homogeneous electric field so that its time-dependence can be modeled by a Dirac's delta function ${\cal H}(t)=-e\vec{E_0}\cdot\vec{r}\;t_0\delta(t)$, where  $e$ is the electron charge  and $t_0$ represents a time unit to compensate the dimensions of the delta function. It can be easily shown that this impulse shifts the momentum of the wavefunctions at $t=0^+$ and, therefore, its effect can be represented by an unitary transformation given by ${\cal U}_{\delta}=\exp(-i t_0e\vec{E_0}\cdot\vec{r}/\hbar)$. From this formula we can deduce that the parameter '$q$' is proportional to the intensity of
the electric field ($q=e\vec{\left|{E_0}\right|}t_0/ \hbar$) and that the vector $\hat{n}$ is parallel to its orientation.

The application of this second transformation yields non-stationary
states $\phi_i^\prime(\vec{r})\chi_s=e^{i q\hat{n}\cdot\vec{r}}\tilde{\phi}_i(\vec{r})\chi_s$
of $\tilde{\cal H}$,  which in turn are eigenstates of ${\cal H}^\prime={\cal U}_d^{\dagger}\tilde{{\cal H}}{\cal U}_d$.
These excited states have a net momentum along $\hat{n}$ with respect to the stationary momentum
distribution of the  eigenstates of $\tilde{\cal H}$. The explicit expression for ${\cal H}^\prime$ reads,
    
\begin{eqnarray}
\label{transham2}
{\cal H}^\prime =\frac{(p_x+q\cos\theta)^2+(p_y+q\sin\theta)^2}{2m^*}+V(x,y) \nonumber \\
    -\alpha^2\frac{m^*}{\hbar^3}L_z\sigma_z +\frac{m^*\alpha^2q}{\hbar^2}\left(\hat{n}_\perp\cdot\vec{r}\right)\sigma_z-\frac{\alpha^2 m^*}{\hbar^2}+o(\alpha^3).
\end{eqnarray}

Due to the shift in momentum space, the kinetic energy is increased according to the effective momentum
$\vec {p^\prime}=\vec{p}+\hbar q \hat{n}$, but the most signicative term in eq. \ref{transham2} is that
given by $(m^*\alpha^2 q/\hbar^2)\left(\hat{n}_\perp\cdot\vec{r}\right)\sigma_z$, where $\hat{n}_\perp\equiv \left(\sin\theta,-\cos\theta\right)$
is a unitary vector perpendicular to $\hat{n}$.  This term can be understood as an effective uniform electric
field which acts perpendicular to the net momentum $\vec{q}=q\hat{n}$ and depends on the spin orientation.
Note that, in this second reference frame (${\cal H}^\prime$), the states
$e^{i q\hat{n}\cdot\vec{r}}\tilde{\phi}_i(\vec{r})\chi_s$ are stationary by construction, i.e.,
they are forced to have a fixed net momentum $\vec{q}=q\hat{n}$. Consequently, it is deduced
that this spin-dependent effective electric field enters in ${\cal H}^\prime$ to fulfill this
constraint and cancel the transverse effect of the spin-orbit interaction.

Coming back to $\tilde{\cal H}$, a particle in a non-stationary state $e^{i q\hat{n}\cdot\vec{r}}\tilde{\phi}_i(\vec{r})\chi_s$ will be deflected
perpendicular to the net momentum  '$\vec{q}$' by a spin-dependent drift term given by $-(m^*\alpha^2q/\hbar^2)\left(\hat{n}_\perp\cdot\vec{r}\right)\sigma_z$,
having opposite sign to that appearing in eq. \ref{transham2}. In fact,
the effective Hamiltonian for the envelope function, defined as

\begin{equation}
\label{heff}
\tilde{\cal H}e^{i q\hat{n}\cdot\vec{r}}\tilde{\phi}_i(\vec{r})\chi_s\equiv e^{i q\hat{n}\cdot\vec{r}} \tilde{\cal H}_{eff}\tilde{\phi}_i(\vec{r})\chi_s ,
\end{equation}

recovers this term with the appropiate sign.

Therefore, the intrinsic spin Hall effect is inherent to the spin-orbit interaction and does not require from a driving
electric field to show its effects.

The present approach can be generalized to include the linear Dresselhaus spin-orbit interaction 
$h_D =(\beta/\hbar)\left(p_x\sigma_x-p_y\sigma_y\right)$ \cite{dress}, where the parameter $\beta$ represents
the strength of this term. In this case, the unitary transformation ${\cal U}_s$ is replaced by

\begin{equation}
\label{uspinc}
{\cal U}_t=\exp\left\{-i\frac{m^*}{\hbar^2}\left[\alpha\left(y\sigma_x-x\sigma_y\right)+\beta\left(x\sigma_x-y\sigma_y\right)\right]\right\}.
\end{equation}

Following the same procedure, a modified effective spin-dependent drift term is obtained

\begin{equation}
\label{hdrift}
h_{drift}=-\frac{m^*\left(\alpha^2-\beta^2\right)q}{\hbar^2}\left(\hat{n}_\perp\cdot\vec{r}\right)\sigma_z.
\end{equation}

Note that  both Dresselhaus and Bychkov-Rashba terms act opposite to each other and, therefore, 
the intrinsic spin Hall effect vanishes in the limit $\alpha=\beta$. This is
in agreement with previous theory since, in this limit, a spin symmetry arises and the orbital motion is decoupled from the spin \cite{schlie}. 
Another interesting consequence derived from eq. \ref{hdrift} is that, up to second order in the strength parameters, the intrinsic spin Hall effect do not distinguish between the two sources of spin-orbit coupling. In general, both terms act as a single effective term having an effective strength $\gamma^2\equiv\mid\alpha^2-\beta^2\mid$ \cite{shen2,sinova2}. However, it would be possible to determine experimentally the dominant spin-orbit interaction if the specific details of the spin flow/spin accumulation were known.

Another relevant point is that this effective spin-dependent drift term  remains unaffected in the presence of the electron-electron interaction since the Coulomb term ${\cal H}_{e-e}=\left(1/4\pi\epsilon\right)\Sigma_{i>j}e^2/r_{ij}$ is invariant under the unitary transformation given by eq. \ref{uspinc} for a system of N particles and is also invariant under translations in momentum space ${\cal U}_d^N=\exp\left[-i q\left(\hat{n}\cdot\Sigma_i\vec{r_i}\right)\right]$. Therefore, manifestations of this transversal spin-dependent drift are expected also in interacting systems \cite{lolo,lolo2}.

Based on the same physical mechanism, a net charge current can be induced 
in the presence of a spin current, i.e., a net flow of spin without charge 
transport. This is known as the inverse spin Hall effect. Looking at eq. \ref{hdrift} it 
can be seen that when the injected momentum 
$\vec{q}=q\hat{n}$ is opposite for each spin eigenvalue, the
effective electric field induced by the spin-orbit interaction points in the same direction for both
spin orientations, thus generating a charge current perpendicular to the spin flow.

In a recent work, the intrinsic inverse spin Hall effect has been observed
in GaAs/AlGaAs multiple quantum wells \cite{invsh}. In this experiment,
focused laser pulses excited valence band electrons  to the conduction band with opposite momenta
for the two spin orientations (aligned along $\hat{z}$) in a small region of the
sample. By means of ultrafast time and space-resolved detection
techniques the dynamics of the electrons were monitored. The time evolution showed 
a charge drift perpendicular to the spin current before any scattering event occured,
proving unambigously the intrinsic character of the observed effect.

In our scheme, the generation of a spin current can be mimicked by the following unitary
transformation

\begin{equation}
\label{uspin}
{\cal U}_i=\exp\left[-i q\left(\hat{n}\cdot\vec{r}\right)\sigma_z\right].
\end{equation}

where the momentum injected to the electrons is opposite for the two $\sigma_z$
spin orientations. Following our approach, the application of this transformation
to the diagonalized Hamiltonian including both Bychkov-Rashba and Dresselhaus
terms (${\cal H}^{\prime\prime}={\cal U}_i^\dagger
\tilde{\cal H}{\cal U}_i$) yields a transverse spin-independent drift term $h_{drift}^{inv}=-\left(m^*q/\hbar^2\right)\left(\alpha^2-\beta^2\right)\left(\hat{n}_\perp\cdot\vec{r}\right)$ .
Therefore, with the constraint of opposite flow for the different spin orientations,
a transverse charge current will be generated, in accordance to the intrinsic inverse
spin Hall effect.

Experimental studies of the spin Hall effect are focused in the
generation/detection of spin currents/spin accumulation in open systems. 
However, since the intrinsic spin Hall effect is inherent to the spin-orbit interaction, it should 
also manifest in closed systems. In what follows, we will show that the spin
Hall effect must show a clear signature in the absorption spectra of closed semiconductor
nanostructures.

\begin{figure}
\includegraphics[width=0.48\textwidth]{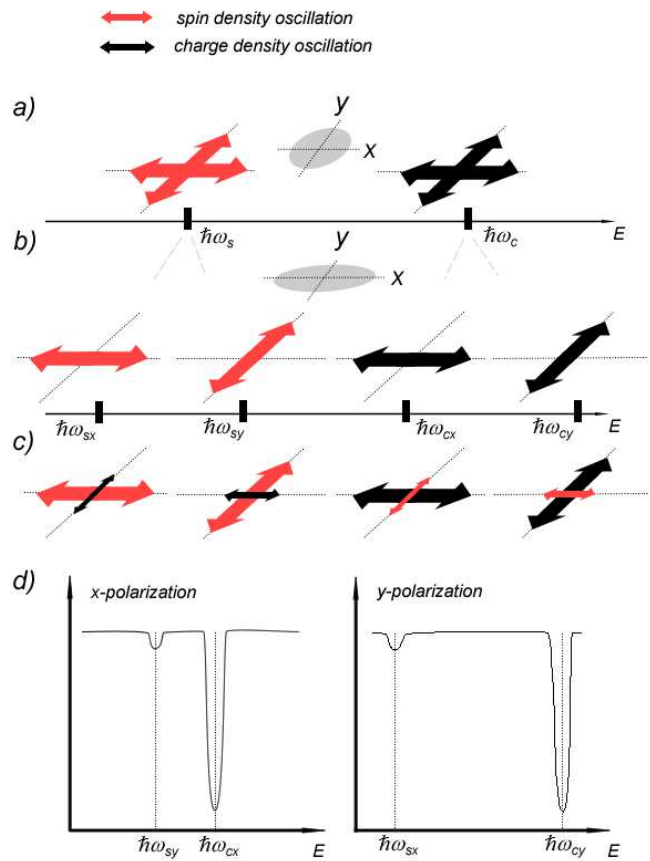}
\caption{ a) Schematic representation of  the charge and spin dipolar oscillation modes in a
symmetric nanostructure. b) Same as panel a) for a deformed nanostructure with broken
degeneracy. c) Same as panel a) for a deformed nanostructure with spin-orbit coupling. The
arrow's size quantifies the character (charge or spin) of the different oscillation modes. 
d) Schematic representation of the far infrared absorption spectra for the different 
directions of polarization of the electromagnetic field.
}
\label{fig1}
\end{figure}

Consider a system of N conduction band electrons confined in a two-dimensional semiconductor
nanostructure. This kind of systems can be excited trhough
intraband transitions lying in an energy range of a few meV \cite{demel}. This energy scale corresponds to
the far infrared region of the electromagnetic spectrum, which is characterized by
wavelengths of the order of the millimeter. Since the typical size of a nanostructure ranges
from a few tens of nanometers to a fraction of a micrometer, the electric field of the 
electromagnetic waves can be safely approximated in the dipolar aproximation, i.e., a uniform
oscillating field across the whole extension of the nanostructure. If the radiation frequency
is resonant, the system can raise to an excited state where the electron density 
oscillates as a dipole, thus generating an oscillating charge current along the polarization direction.
Apart from these dipolar charge modes, these systems exhibit dipolar spin modes where the
spin density oscillates as a dipole in a given direction. This modes are characterized by a lower energy
than the dipolar charge ones \cite{toni,lolo}.

In panel a) of Fig. 1 there are represented schematically the frequencies of the different
oscillation modes corresponding to a symmetric nanostructure without spin-orbit interaction \cite{toni,lolo}.
In this case, both charge ($\omega_c$) and spin dipolar modes ($\omega_s$) are degenerate, since
the excitation is equivalent for all the polarization directions.
The degeneration disappears when the nanostructure is deformed, resulting in the fragmentation of
the modes due to the anisotropy \cite{lolo2}. In panel b) of Fig. 1, it is represented schematically this situation
where the dipolar charge mode splits into two modes corresponding to the charge oscillations in the
main axes of the nanostructure. The higher frequency mode ($\omega_{cy}$) corresponds to the axis
of higher confinement, while the lower one ($\omega_{cx}$)  corresponds to the less confined one.
Similarly, the spin dipolar mode splits into $\omega_{sy}$ and $\omega_{sx}$  where $\omega_{sy}>\omega_{sx}$.

When the spin-orbit interaction is included the character of the different modes is hybridized and we no
longer have pure charge or pure spin dipolar modes. Due to the intrinsic spin Hall effect, the
dipolar oscillation of the charge modes will induce a transversal oscillation of the spin density \cite{lolo,lolo2}. 
At the same time, due to the inverse spin Hall effect, the oscillation of the spin density corresponding to the
spin dipolar modes will induce a charge oscillation in the transverse direction. In panel c) of Fig. 1, the hybridization
of the different modes is represented schematically. Each oscillation mode has a dominant charge(spin) character
plus a small transversal spin(charge) contribution which are represented by arrows of different size.

This particular distribution of the spin and charge oscillation patterns through the different modes has a
clear signature in the far infrared absorption spectrum. For instance, a system irradiated with light
polarized along the x-axis will show a dominant absorption peaks at $\omega_{cx}$ and a less pronounced one at
$\omega_{sy}$ (Fig. 1 panel d), left). On contary, a system irradiated with y-polarized light will show its
dominant peak at $\omega_{cy}$ and a smaller one at $\omega_{sx}$ (Fig. 1 panel d), right). This complementarity
of the absorption spectra for the different directions of polarization, i.e. closest absorption peaks for polarization in 
the axis of lower confinement and separated absorption peaks for polarization in the axis of higher confinement,
 is a unambigous proof of the intrinsic spin Hall effect since it demonstrates the spatial correlation between a charge dipolar mode
and the corresponding spin dipolar mode in the transverse direction.

In summary, we have analytically shown that the spin-orbit interaction can act as a transverse
effective uniform electric field whose sign depends on the orientation of the out-of-plane spin. The effect is inherent
 to the spin-orbit interaction and doesn't require any driving electric field, so it can manifest in both open and closed systems. 
We have proposed an experiment to observe the signatures of the intrinsic spin Hall effect in the far infrared 
absorption of a deformed  semiconductor nanostructure where a clear signature must be observed.

\end{document}